# Aerogel from sustainably grown bacterial cellulose pellicle as thermally insulative film for building envelope

*Blaise Fleury[1,‡], Eldho Abraham[1,‡], Joshua A. De La Cruz[2], Varun S. Chandrasekar[1], Bohdan Senyuk[1], Qingkun Liu[1,†], Vladyslav Cherpak[1], Sungoh Park[1], Jan Bart ten Hove[1], and Ivan I. Smalyukh[1,2,3,4,*]*

[1]Department of Physics, University of Colorado, Boulder, Colorado 80309, United States

[2]Materials Science and Engineering Program, University of Colorado, Boulder, Colorado 80309, United States

[3]Soft Materials Research Center and Department of Electrical, Computer and Energy Engineering, University of Colorado, Boulder, Colorado 80309, United States

[4]Renewable and Sustainable Energy Institute, National Renewable Energy Laboratory and University of Colorado, Boulder, Colorado 80303, United States

Keywords: cellulose aerogel, bacterial cellulose, insulation, building envelope, thermal conductivity, sustainable

**Abstract**

Improving building energy performance requires the development of new highly insulative materials. An affordable retrofitting solution comprising a thin film could improve the resistance to heat flow in both residential and commercial buildings and reduce overall energy consumption. Here we propose cellulose aerogel films formed from pellicles produced by the bacteria *Gluconacetobacter hansenii* as insulation materials. We studied the impact of density and nanostructure on the aerogels' thermal properties. Thermal conductivity as low as 13 mW/(K·m) was measured for native pellicle-based aerogels dried as-is with minimal post-treatment. The use of waste from the beer brewing industry as a solution to grow the pellicle maintained the cellulose yield obtained with standard Hestrin–Schramm medium, making our product more affordable and sustainable. In the future, our work can be extended through further diversification of the sources of substrate among food wastes, facilitating larger potential production and applications.

**Introduction**

As the human population along with the demand for energy keep growing, the need for reducing energy consumption similarly increases. Residential and commercial building energy consumption amounts to ~40% of total energy consumption in the USA and ~37% in the EU.[1] The energy spent on heating and cooling buildings across the world is massive and should be reduced in the near future. Simply relying on regulations for new buildings will not cope fast enough with the present situation. Improving the overall insulative properties of existing edifices becomes mandatory to achieve such energy reduction goals. Understanding these issues, in 2014 the United States Department of Energy (DOE) issued a roadmap for emerging technologies regarding building envelope research and development, including new building construction and retrofitting of existing buildings.[2] Therein their performance target for retrofit products is to achieve a thermal resistance equivalent to an R-value of 12 per inch of material by 2025.

To address this goal, we investigated the fabrication of cellulose aerogels as a candidate to provide the required insulation in a product that is both sustainable and environmentally friendly. Cellulose is considered a green product as it is widely available, renewable, and is commonly extracted from wood pulp, cotton or grown by bacteria.[3,4,5] Most recent applications of cellulose include 3D printed metamaterials[6] and MOFs,[7,8] biomedical applications due to its biocompatibility[9,10] and thermal insulative properties.[11,12] Even with an increasing research interest in cellulose, most studies involve many processing steps such as cellulose extraction, purification, and post-treatments to obtain purified nanocellulose before fabricating a product.[13,14] A considerable interest was found in drying directly a bacterial cellulose pellicle into an aerogel after a few steps of solvent exchanges.[15] This method streamlines the fabrication of cellulosic aerogel. Despite this very attractive idea, the ensuing material was so far not bound to practical applications and remained cost prohibitive for applications like building insulation due to the use of expensive growth medium. Recent work compiled studies of alternative carbon source and nutrients used to grow cellulose.[16] In particular, the use of beer brewing waste was demonstrated as an affordable alternative growth solution with high yield, though characterization of the ensuing materials did not focus on properties related to thermal insulation of building envelopes.[17]

Here we focused on developing cellulose aerogel insulative materials, directly produced by bacteria, exhibiting maximal thermal resistivity while minimizing processing steps. We used waste from the beer industry to reduce the production cost of the aerogel while also optimizing parameters to maximize the thermal barrier properties. We explored the use of additives to tweak *in situ* the nanostructure of the aerogels. We thoroughly characterized the thermal properties of our samples and found them to satisfy the DOE's next generation building envelope insulation materials target of 12 per inch R-values.[2] In addition, our fabrication approach is poised to both reduce production costs and lower the environmental impact of our technology by using waste as a nutrient source for the bacteria.

**Materials and Methods**

Standard Hestrin–Schramm (*HS*) medium was prepared as described in the literature.[18] Briefly, the reference substrate final solution consisted of glucose (20 g/L), peptone (5 g/L), yeast extract (5 g/L), disodium phosphate (2.7 g/L), and citric acid (1.15 g/L). The pH of the HS solution was 6. Beer waste (BW) derived growth solution was made from beer waste (fermented or not) obtained from local breweries in Boulder, Colorado. The acidic waste liquid is brown and contains sugars, proteins, flavor molecules[19] such as acids and esters, various salts and fibers adding up to ~70 g/L of solid content. The exact composition varies somewhat depending on the exact waste used, but this did not seem to affect the results in terms of yield or properties of aerogel materials. After all, this waste is always based on malt and hop extracts. A few weight percent of ethanol can be present when fermented waste is used, which is consistent with previous work.[17]

When preparing our initial source medium, the pH was adjusted to 5.5 by adding 2M NaOH solution prior to autoclaving the solution for 30 minutes at 121 °C. Carboxymethyl cellulose (CMC) (15 g/L) was added to the growth medium before sterilization to yield thinner cellulose fibers. Once cooled down to room temperature, glucose was added to the solution immediately before inoculation with the bacteria. Different concentrations were tested. Optimal final glucose concentration in BW based solution was found at 10 g/L. For both the BW and HS media, the sugar solution was autoclaved separately under the same conditions and combined with the rest of the growth medium prior to use.

*Gluconacetobacter hansenii* (ATCC® 53582™) was purchased from ATCC and revived according to their suggested protocol. This particular bacterial strain has been selected as it is known to swiftly produce homogeneous cellulose pellicle.[20] The sample was rehydrated and transferred to a culture tube with 5 mL of HS growth medium. The solution was kept in an incubator (26 °C and 250 rpm) for 5 days. The solution became slightly hazy, indicating successful bacteria revival. Agar culture plates were inoculated with the revived *G. hansenii* and incubated for 4 days. Single colonies were then collected and allowed to multiply for 5 more days in a 5 mL culture tube in the incubator (26 °C and 250 rpm). The 5 mL solution was then injected in one liter of either HS substrate or BW-based growth solution.

Cellulose production by *G. hansenii* is visible after a few days of incubation (26 °C static conditions) as a solid floating at the liquid air interface (Figure 1 a-d). Aerogel production from cellulose pellicles follows steps similar to previous literature reports[15] and is described as follows. Pellicles are typically harvested after three weeks of growth (Figure 1h). The thickness of the pellicle varies between 1 and 20 mm depending on nutrient concentration, the aspect ratio of the container, and the duration of the growth. To accelerate the washing step of fabrication, the cellulose is cut into smaller pieces and washed with 1% NaOH at 80°C for 1 hour (Figure 1e). Then the pellicles are washed with water until the pH remains neutral (Figure 1f). We explored the scaling up of the production of cellulose by increasing both the size and the number of pellicles produced at the same time (Figure 1 g-j). Up to 10 cellulose pellicles ~0.1 m² (10x15 inches²) were produced in the laboratory at a given time.

Slowly, the cellulosic hydrogel's water is exchanged with ethanol by pure immersion. The solvent exchange is repeated five times at room temperature, with each step lasting for at least 6 hours. The cellulosic alcogel is then dried via super critical $CO_2$ using a Tousimis Automegasamdri 915B critical point dryer. The chamber is filled with ethanol before placing the alcogel. After closing the lid, the system is cooled down and liquid $CO_2$ is injected. The ethanol present in the chamber is pushed out through the purge line as more $CO_2$ is added. Once most the ethanol is purged, the machine is heated up to 40°C, yielding a pressure of 1300 PSI (8.96 MPa). Under these conditions, the $CO_2$ is supercritical. Remaining ethanol within the sample is exchanged with supercritical $CO_2$ and is collected at the bottom of the chamber due to an increased density. The system is left for several hours before releasing slowly the supercritical fluid (25 PSI/minute, 0.17 MPa/minute) until the chamber is at atmospheric pressure. The cellulose is now an aerogel with a density ranging from ~6 to 28 mg/mL, depending on the growth conditions (duration, glucose concentration, use of CMC additive, etc.). To increase the density of the cellulosic aerogel, we plastically deformed the aerogels via uniaxial compression. Squeezing of the samples was performed uniformly across the sample using two wooden or plastic non-stick slabs covering the entire area of the sample before applying pressure. Using this method densities up to 40 mg/mL were obtained.

Thermal conductivity $k$ of aerogel films was measured using a comparative method,[21] wherein the same heat flow is passed through a stack of aerogel and reference films of comparable area with the reference film facing the heat source and aerogel film facing the heat sink. Thus, by measuring the difference of temperatures at both surfaces of the films, thermal conductivity $k$ was found as $k=k_0(\Delta T_0/\Delta T)(A_0/A)(d/d_0)$, where $k_0$ is thermal conductivity of a reference film, $d$ and $d_0$ are the respective thickness of the sample and reference, $A$ and $A_0$ are the respective cross-section area of the sample and reference ($A_0/A \approx 1$), and $\Delta T$ and $\Delta T_0$ are differences between temperature at two surfaces of the sample and reference, respectively. As a reference we used an aerogel with comparably low thermal conductivity $k_0=14$ mW/(K.m) as confirmed by measurements using guarded hot plate method in certified characterization facilities. The temperature of each film's surface was measured using FluxTeq thermocouples and acquisition system.

**Instrumentation**

Thermogravimetric analysis (TGA) was performed in $N_2$ atmosphere at 25-500 °C. TGA runs were performed in a Netsch STA 449 F1 Jupiter thermogravimeter with alumina crucible at a heating rate of 10 °C/min. Thermal stability of the sample was calculated using a basic mass loss rate expression $dm/dt$ and normalized to the total quantity of mass lost.

Scanning Electron Microscopy (SEM) images were taken on a FEI Quanta 600 SEM at the National Renewable Energy Laboratory (NREL). The sample were sputtered with a thin layer (<~5 nm) of Pt/Au mix and imaged under 3 kV accelerating voltage.

Fourier-Transform InfraRed (FTIR) spectroscopy experiments were performed in the mid-infrared (4000–500 cm$^{-1}$) using a Nicolet 750 Magna FTIR spectrometer equipped with a Pike brand integrating sphere with a KBr beam splitter and an MCT/A detector.

UV-visible measurements were done on a Cary 500 Scan UV-Vis-NIR spectrometer equipped with a Labsphere DRA-CA-5500 integrating sphere.

Mechanical measurements were made using a DMA 850 from TA instruments with a standard tension clamp attached. Applied force and displacement of the clamp were recorded. The initial dimensions of the sample were used to convert the measurements to stress and strain using TRIOS software.

**Results and Discussion**

Bacterial cellulose pellicles produced by *G. hansenii* cultured in HS medium or BW-based substrate look similar. The yield of cellulose production per liter of culture medium after three weeks in HS medium is 7 g/L while it is only 5.5 g/L for the BW as received. The addition of 10 g/L of glucose to the BW increased the yield to 6.5 g/L. Further increase to 20 g/L of added glucose to the BW solution did not increase the cellulose yield. These results are in agreement with a previous research on the production of cellulose using beer waste with addition of glucose.[17] From this, BW-based beer production waste medium appears to be a liable, economical alternative to HS medium for growing cellulose pellicles as it can be obtained free of charge.[22]

The bacteria studied directly excrete cellulose.[23] Several channels on their membrane produce a thin thread of cellulose nanofibril. These nanofibrils aggregate into thicker ribbons due to hydrogen bonding (Figure 2a). When CMC is added to growth solutions, the aggregation of elementary fibrils is inhibited as the CMC molecules intercalate between cellulose fibrils (Figure 2b), prohibiting hydrogen bonding.[24] The native production of cellulose by *G. hansenii* was imaged via dark field optical microscopy (Figure 2c). However, in presence of CMC, the thinner threads of bacterial nanocellulose could not be optically resolved via dark-field imaging (Figure 2d). As the bacteria were moving under the microscope and their primary motion mechanism requires the production of cellulose,[25,26] we can confirm the production of cellulose with a reduced diameter due to CMC's addition. Cellulose hydrogel pellicles obtained with CMC additive are transparent (Figure 2e), consistent with the reduced light scattering and smaller diameters of its constituent nanofibers.

The transparency of cellulose pellicles grown in the presence of CMC is directly linked to the thinner cellulose constituting the pellicle. Furthermore, the pellicle density was lower when CMC was present. Regardless, the transparency did not persist once the pellicle was dried (Figure 3). Under optimal conditions, supercritical drying prevents hornification and shrinkage of the pellicle. The loss of transparency here is mainly explained by an increase of refractive index contrast between the cellulose material and air as compared to the much smaller contrast in the case of hydrogels. Three transparent pellicles harvested at different times yield different densities once dried (7.2, 9.4 and 11 mg/mL). SEM imaging of the aerogels shows the disordered cellulose fiber network whose diameter are in the tens of nanometer and can be as small as ~10 nm (Figures 3g, h, and i). These dimensions are in agreement with previous work using CMC[27] and are much smaller than within the native cellulosic aerogel.[15] The absence of bundled fibers on the micrographs also confirms the very limited extent of hornification, if any, during the drying process. On the contrary, if the hornification is of interest, the use of less polar solvents as intermediates, instead of ethanol, could be a lever to control the bundling of the nanofibers and thus allow tuning of the light scattering properties of the final sample.

Qualitative thermal characterization of our aerogels is shown on Figures 4a, 4b. The cellulose aerogel is placed over a hot surface (72°C) and both traditional photography and infrared (IR) thermographs are shown. On the IR image (Figure 4b), the film remains blue (58°C) while the hot backplate is pink, showing the relatively large difference in temperature between the two surfaces. Thermal stability of the film was assessed using thermogravimetric analysis (Figure 4c). The mass stays stable with temperature up to ~230 °C where a drop is observed as the cellulose starts to degrade. The measured weight loss rate is maximal at 321 °C. These results are comparable to published literature data, as expected for pure cellulose sample.[28] An outgassing pre-treatment at 60 °C was performed before recording the thermogravimetric data. During that process, moisture and adsorbed molecules are removed from the sample. Up to ~8-10% of the total mass is removed with degassing, consistently with previous work.[29] The outgassing pre-treatment at 60 °C is an advised procedure for to maximize thermal insulation and allow better durability of a product. It was utilized before each thermal conductivity characterization of our samples reported in this work. Quantitative analysis of the thermal properties of the films was made using heat flux sensors (see experimental section for details). The thermal conductivity (Figure 4d) and the R-value per inch of material (Figure 4e) are given as a function of the measured density of the cellulose aerogel. The thermal conductivity includes contributions from gas conduction within the pores and across the networked cellulose fibers and emissivity related contributions (we note that the contribution due to thermal-range emissivity can be disregarded for these materials at ambient temperatures). When the porosity is too high, the thermal transfers through the gas conduction within the pores are significant, yielding a higher thermal conductivity. In that sense, smaller pores are preferred. However, when reducing the pore size by increasing the density of fibers, the latter's contribution to the thermal conductivity increases. It follows that minimum thermal conductivity is achieved when the competing factors of both pore size and density of fibers are optimized. The optimum density is found at densities around 25 mg/mL, where the conductivity could be assessed to be as low as k~ 12-13 mW/Km. The corresponding R-value per inch is close to 12, meeting the DOE's 2025 performance target table for emerging building envelope technologies.[2] The lowest measured value obtained was 13 mW/(K·m) for a sample whose density was measured to be 28 mg/mL. To the authors' knowledge, this is the lowest thermal conductivity reported for a cellulose aerogel.[12,30,31] Standalone, these aerogels already present excellent insulation properties, but they would further outperform when embedded in multi-layer insulation blankets (MLI).[32,33] Briefly, within this envisaged embodiment of a building insulation material, our cellulose aerogel would be encapsulated between radiation reflective materials, such as a metal foil, or films with metallized coatings to reduce the radiant heat transfer of our system. The system can be more efficient when multiple alternating layers are used. Although demonstrating this is outside of this present work, we hypothesize that, overall, the thermal insulation would be enhanced, and the cellulose aerogel would be protected in this implementation. In particular, fire protection of the cellulose aerogel is a major concern and could be mitigated through encapsulation with specific films, such as multi-layer polymer metal laminates.[34]

Cellulose aerogels can easily be handled by hand and are quite flexible (Figure 5a). Their mechanical properties under tension were characterized using a DMA 850 from TA instruments with a standard clamp (Figure 5b). Rectangular shape samples, ~6 mm wide and ~10-20 mm long, were cut using a razor blade. Characteristic stress versus strain curves are given for a native cellulose aerogel (Figure 5c) and a CMC modified cellulose aerogel (Figure 5d). All curves show an initial regime were the fibers are aligned along the pulling direction. Initial Young's modulus

E can be as low as 10-20 kPa. Then the slope steepens as the fibers are pulled (E~100-150 kPa). The randomness of the cellulose network creates a variety of response and maximum elongation before failure up to >30% elongation was measured. Cellulose aerogel with CMC additive consists of thinner fibers so they are less stiff than native ones (E~50 kPa for the second regime). Even though cellulose aerogels are not tough materials, these results show they can be handled without risking film tearing, confirming the potential of bacterial cellulose as insulation material for building envelope. In addition, they are easily cut to any size and shape with a razor blade or a utility knife.

The white appearance of our cellulose aerogel is due to strong Mie scattering by its porous network structure. When the cellulose aerogel density is sufficiently low, ~20% of the visible light can be transmitted through as scattered transmission (Figure 6a) while it maintains a low thermal conductivity, similar to or lower than air. Sandwiching such cellulose aerogels between glass panes in IGU can provide a diffuse light much appreciable in applications such as sky lights, daylighting and in various privacy windows while reducing solar heat gain and maintaining superior thermal insulation. The window will not heat up as the light is not absorbed but rather scattered. Looking beyond hot climates, this solution could be also useful for greenhouses where light needs to be spread equally on all the plants.[35] FTIR spectrum shown in Figure 6b is typical for cellulose-based materials. The total absorption around 1000 cm$^{-1}$ indicates the porous material would absorb mid-IR from black body emission of objects at room temperature. Then, because of the low thermal conductivity of the aerogel, the energy would be reemitted and reabsorbed multiple times but effectively reflected backwards rather than penetrating through the aerogel film thanks to the aerogel's porous network. Overall, the transmission of this mid-IR energy linked to the radiation of indoor object is expected to be low, and most of it should be reflected towards the inside of the building, though the metalized coatings and metal foils briefly discussed above can be used to further reduce the emissivity contribution to the thermal barrier behavior of our materials.

**Conclusions**

In this work, we report the bacterial production of cellulose pellicles as an insulative material for building envelope applications. *G. hansenii* natively produce cellulose as part of their motility process. When grown in large colonies, the bacteria weave in concert a porous network of intertwined cellulose fibers, forming a nanoporous extracellular matrix. This cellulose pellicle, once harvested and dried with supercritical $CO_2$, results in an aerogel. A considerably low thermal conductivity of 13 mW/(K·m) was measured on native cellulose aerogels. This performance shows our material is suitable for next generation thermal barriers for building envelopes. We also modified our pellicles using CMC additive during the growth to keep the cellulose fibers thin, which could be of interest to control different properties of aerogel materials depending on the needs of specific applications. For example, the increased accessible area obtained with the thinner fibers could become useful in imparting hydrophobicity of aerogels through various surface modifications. One extra advantage of that CMC-enabled aerogel material is it can be cut to size but also sewn into larger pieces. Mechanical and thermal gravimetric characterization confirmed the real potential of this material for the building envelope insulation industry. In particular, cellulose aerogels could be used in multilayer insulation blankets to not only further increase the thermal insulation of the material but also add fireproofing properties. We showed a substrate medium based on waste from brewing beer yields a cellulose output similar to relatively expensive HS medium while also allowing to optimize mesoscale structure and tune density needed to maximize thermal barrier properties. This strategy based on recycling waste not only reduces the cost but makes the production sustainable. We expect to see other sources of waste, especially from the food industry, to perform as well as the beer waste we used for this research, which is one future research avenues to be pursued.

**Corresponding Author**

*ivan.smalyukh@colorado.edu

**Present Addresses**

†Laboratory of Atomic and Solid State Physics (LASSP), Department of Physics, Cornell University, Ithaca, NY 14853, United States

**Author Contributions**

The manuscript was written through contributions of all authors. All authors have given approval to the final version of the manuscript. ‡These authors contributed equally.

**Funding Sources**

This work was supported by the U.S. Department of Energy, ARPA-E award DE-AR0000743

**Acknowledgments**

We would like to thank Jeffrey C. Cameron, Allister W. Frazier, Evan Johnson, and Karthik Peddireddy for fruitful discussions and various technical assistance. We thank Asher Brewing Co. for providing us with the beer waste. We also thank Jao van de Lagemaat for the access to the SEM facility at NREL.

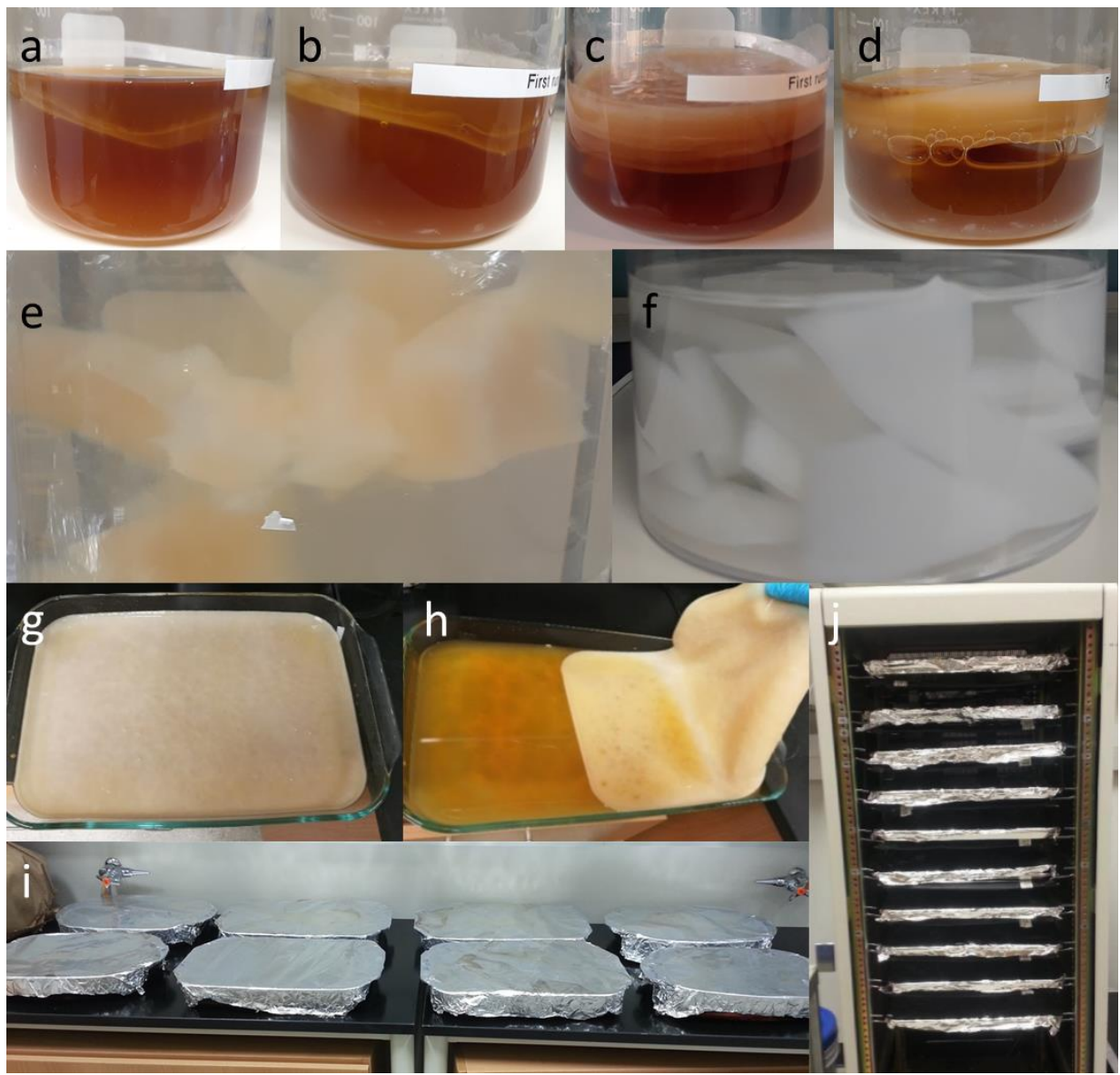

**Figure 1.** Production of a cellulose pellicle by bacteria using waste from the beer industry as a growth medium. First row: pictures of the growing pellicle close to the surface after 3 days (a), 7 days (b), 10 days (c) and 14 days (d). Second row: harvested pellicle is cut into pieces and washed first with 1% NaOH at 80 °C (e) then with water (f). Scaling up the cellulose production using larger pans (g-j). Pictures of 0.1 m² fully grown cellulose pellicles (g and h). Picture of 8 pans prepared for the cellulose production (i). Vertical racking of the pans to save space (j).

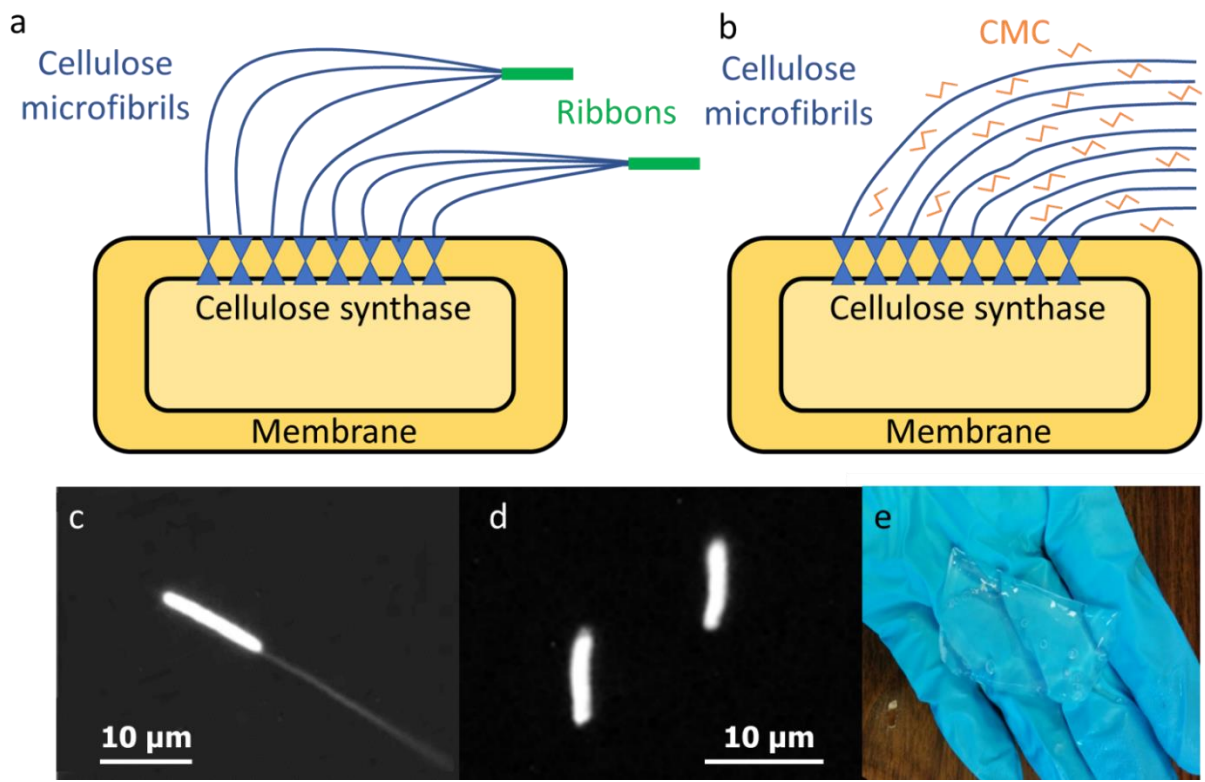

**Figure 2.** Schematics of cellulose production by a bacterium: native cellulose (a) and cellulose produced with CMC additive (b). In absence of CMC, the microfibrils aggregate into ribbons, which can also further aggregate. CMC molecules intercalate between elementary fibrils through hydrogen bonding, preventing the aggregation of fibrils into ribbons. Dark-field microscopy images of bacteria producing cellulose without (c) and with CMC (d). The thinner thread of cellulose could not be optically resolved (d). Picture of a transparent cellulose pellicle obtained with CMC additive (e).

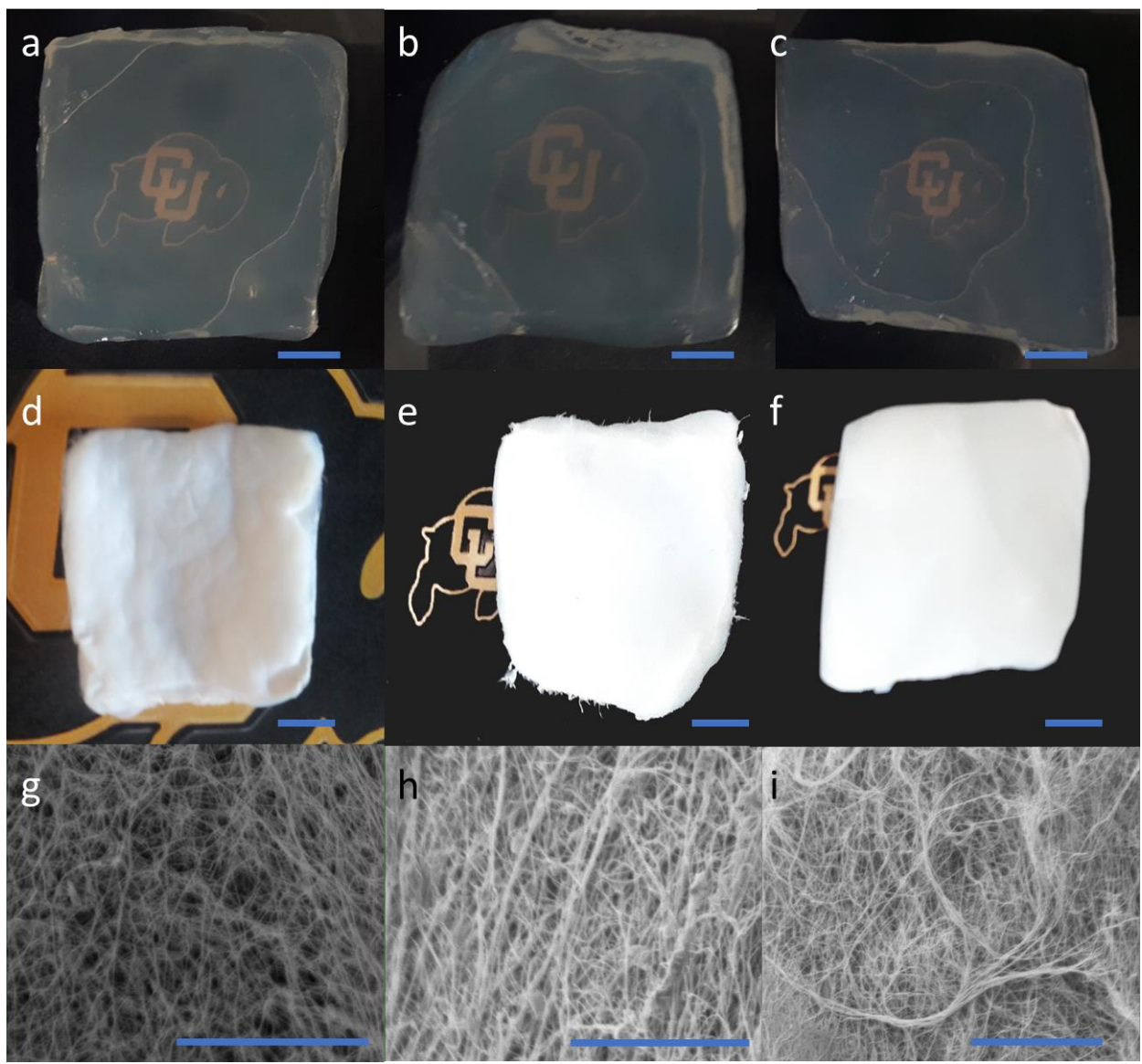

**Figure 3.** Cellulose pellicles grown with 1.5 wt% of CMC additive yielding different densities once dried, depending on when the bacteria cellulose was harvested. (First column: 7.2 mg/mL, second column: 9.4 mg/mL and third column: 11 mg/mL). The wet pellicles (a,b,c) are transparent due to thinner cellulose fibers and refractive index matching with water. Scale bar is 1 cm. After drying, the aerogels (d,e,f) are opaque, flexible and (in lateral dimensions) measure ~5×5 cm². Scale bar is 1 cm. Corresponding SEM images (g,h,i) of cellulose aerogels (d,e,f) show a reduced thickness of the fibers due to the addition of CMC during the growth. Scale bar is 5 μm.

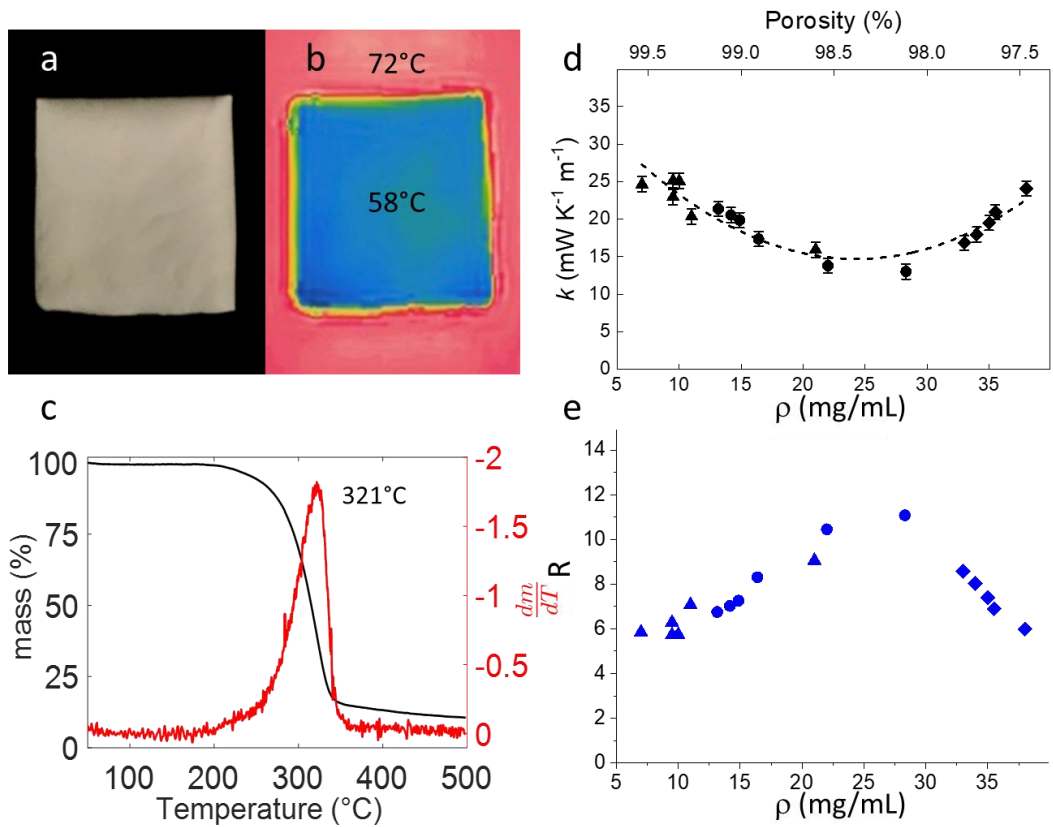

**Figure 4.** Conventional photograph of a ~4*5 cm² cellulose aerogel (a) and infrared image of the aerogel (blue) on top of a hot surface (pink, 72°C) (b). The blue color indicates the surface of the film stays colder (58°C), qualitatively demonstrating the thermal insulative properties of the material. Thermogravimetric analysis in c shows thermal stability of the cellulose up to 230 °C and a maximum dm/dT at 321°C. The cellulose aerogel characterized here was heated to remove moisture (~8-10% of the initial weight) before performing the TGA experiments. Quantitative characterization allowed to measure the thermal conductivity (d) and the R-value per inch of material (e) versus the density or the porosity of the cellulose aerogel. Solid circles and triangles show respectively the characteristics of cellulose without and with CMC. Solid diamonds show properties of samples whose density was increased by uniformly squeezing cellulose samples without CMC.

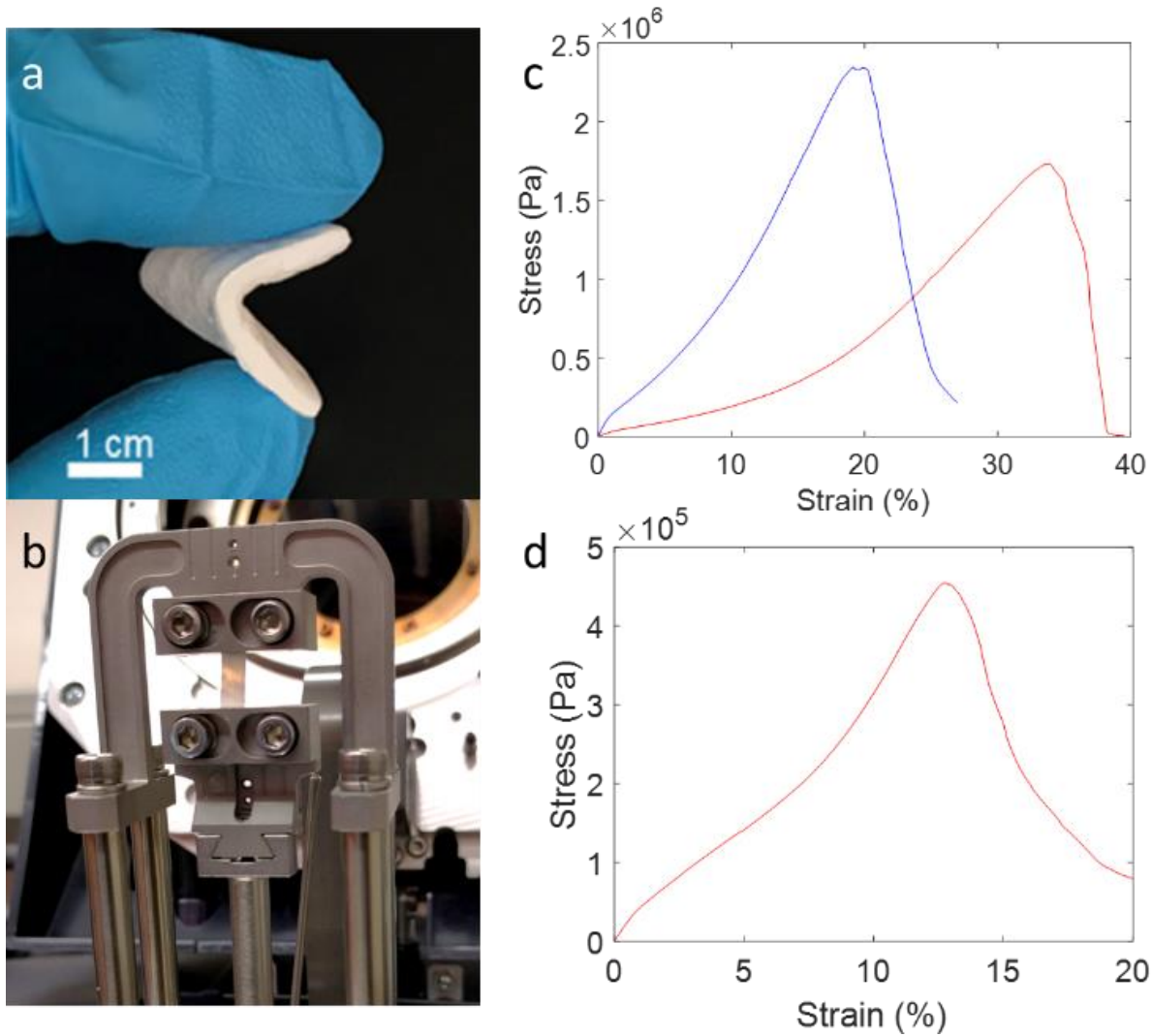

**Figure 5.** a) Picture showing qualitatively the flexibility of a cellulose aerogel. b) Picture of the device used for the mechanical analysis (DMA850 from TA). A cellulose aerogel is attached to the test clamp and locally appears translucent as it is being thinned and torn upon tension. c) Typical examples of standard pull results for a regular cellulose aerogel. During the measurement, the cellulose fibers are first aligned, then they are elongated. Finally, they are torn one after the other without showing a catastrophic failure. The blue curve corresponds to a sample where the fibers were initially more aligned than in the sample corresponding to the red curve. d) Measurement for a cellulose aerogel with CMC additive of similar density shows a weaker response while maintaining a similar maximal elongation.

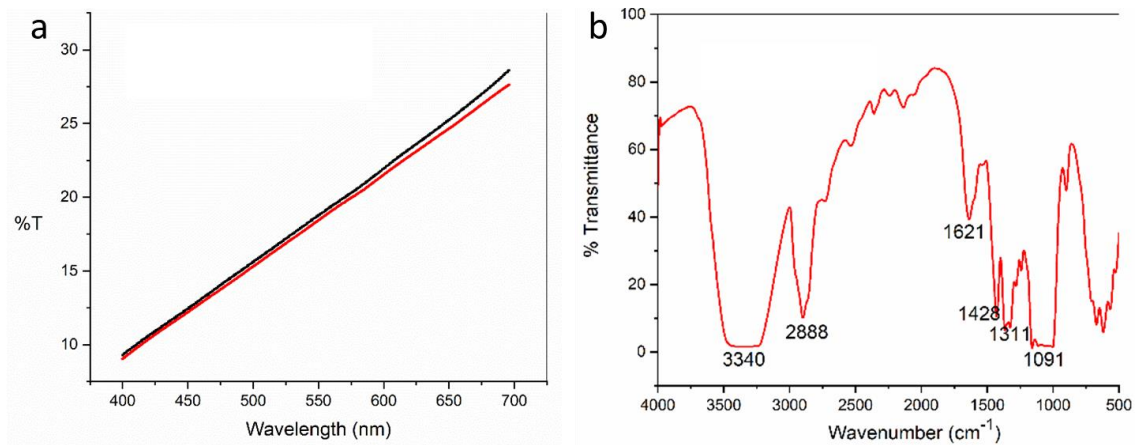

**Figure 6.** a) Low-density cellulose aerogel visible transmission spectra showing ~20% of scattered transmission at 550 nm. Both the total transmission (solid black curve) and scattered transmission (solid red curve) are plotted. b) FTIR measurement of transmittance of a low-density cellulose aerogel.

TOC

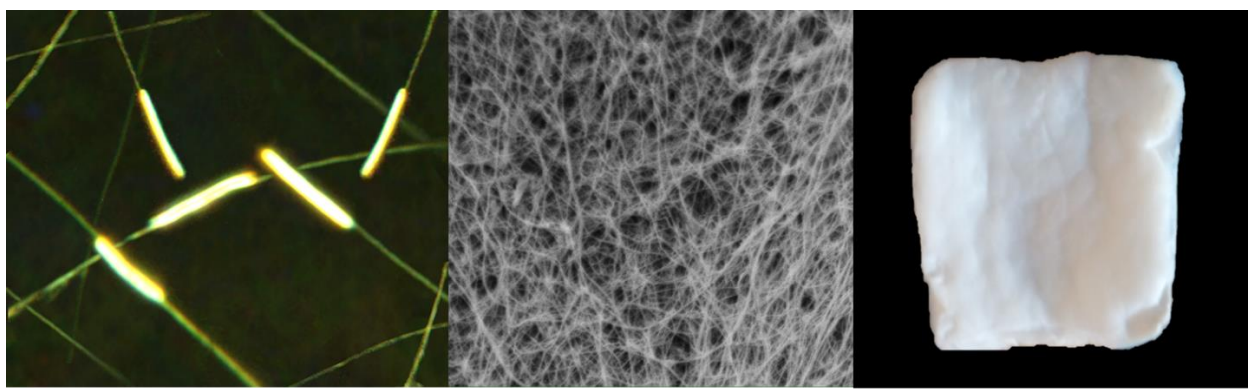

Bacteria fed with waste weave cellulose into a nanostructured thermal insulator